\documentclass[11pt]{article}
\usepackage{booktabs}
\usepackage{multirow}
\usepackage[final]{acl}

\usepackage{times}
\usepackage{minted}
\usepackage{stfloats}
\usepackage{dpfloat}
\usepackage{latexsym}
\usepackage{subcaption} 
\usepackage{amsmath}
\usepackage{amssymb}
\usepackage{wrapfig}
\usepackage{subcaption}
\usepackage[most]{tcolorbox}
\newtcolorbox{rqbox}{
    colback=gray!10,
    colframe=gray!60,
    boxrule=0.5pt,
    arc=3pt,
    left=6pt,
    right=6pt,
    top=6pt,
    bottom=6pt
}
\usepackage[T1]{fontenc}

\usepackage[utf8]{inputenc}

\usepackage{microtype}

\usepackage{inconsolata}

\usepackage{graphicx}

\newcommand{\mypara}[1]{\noindent\textbf{#1.}}

\newcommand\blfootnote[1]{%
  \begingroup
  \renewcommand\thefootnote{}\footnote{#1}%
  \addtocounter{footnote}{-1}%
  \endgroup
}

\title{\textit{When Safety Routing Breaks}: \\Understanding Alignment Fragility under Benign Fine-Tuning}

\author{
  \textbf{Yitong Guo\textsuperscript{1}}\footnotemark[1]\quad
  \textbf{Xiaoyi Chen\textsuperscript{1}}\footnotemark[1] \quad
  \textbf{Siyuan Zhang\textsuperscript{2}}
\\
  \textbf{XiaoFeng Wang\textsuperscript{3}}\quad
  \textbf{Haixu Tang\textsuperscript{1}}
\\
  \textsuperscript{1}Indiana University Bloomington\quad
  \textsuperscript{2}Tsinghua University\quad
  \textsuperscript{3}Nanyang Technological University
\\
  \small{
    \textsuperscript{*}Equal contribution. \quad
    \textbf{Corresponding Author:} Xiaoyi Chen (\href{mailto:chxiaoyi@iu.edu}{chxiaoyi@iu.edu})
  }
}

\begin{document}
\maketitle
\begin{abstract}
Benign fine-tuning severely weakens the safety alignment of large language models (LLMs), so we study why refusal behavior is so fragile.  
While prior work often attributes this failure to gradient conflict, we propose a fundamentally different Fisher-geometric explanation: safety Fisher is low-rank, and alignment makes the safety geometry flatter while preserving an output-routing pathway. 
After 100 benign fine-tuning examples, this pathway is selectively re-sharpened in output-side MLP modules, explaining the asymmetric fragility: safety can collapse to high attack success rates, while general utility degrades mildly.

The routing view also explains why few safety examples can restore refusal behavior, indicating that internal safety-relevant representations are preserved.  
Finally, we show that LoRA and ASAM mitigate early collapse by suppressing output-side sharpness, but their protection weakens at larger fine-tuning scales.  
Overall, safety failure is best understood as a disruption of a low-rank output-routing mechanism.
\blfootnote{Code:\url{https://github.com/Godblessmycode1/safety_routing}}
\end{abstract}

\section{Introduction}
LLMs are routinely adapted after alignment to improve performance on downstream utility tasks. Such adaptation is often benign: the fine-tuning data may consist of domain-specific question answering, coding, or reasoning examples, with no harmful instructions. Ideally, this process should preserve the safety behavior. Yet recent work has shown that even benign fine-tuning can substantially weaken refusal behavior and increase attack success rate \citep{qi2024fine, zhan2024removing}. This raises a basic question: why is safety alignment so easy to break?

A natural explanation is gradient conflict: utility fine-tuning may update parameters in directions that interfere with safety alignment~\cite{guan2025benign,he2024your}. 
Prior work manipulates and carefully selects outlier benign samples to break safety.
However, this explanation alone does not capture the empirical pattern we observe. In our experiments, we find that low-conflict and random sample subsets all cause comparable safety collapse, ruling out gradient conflict as the primary cause.

We argue that this fragility arises because \emph{safety alignment primarily controls how harmful representations are routed to the output}.
During pretraining, LLMs acquire both general knowledge and representations of harmful content, which coexist in the same parameter space. Post-training alignment actually solves a routing problem: the model should map the harmful representations to refusal behavior instead of compliant answers. 

Fisher measurements support this routing view. First, safety is more concentrated than utility. On the baseline model, the safety Fisher has a larger top-1 concentration than the utility Fisher (\(0.491\) vs.\ \(0.271\)), with the concentration most pronounced in the final layers (\(0.718\) vs.\ \(0.223\)).  
Second, alignment compresses the safety Fisher, reducing layer-wise top
eigenvalues by roughly two orders of magnitude across most layers.
These findings suggest that alignment makes the safety geometry
flatter while preserving a low-rank routing pathway from safety-relevant
internal states to refusal behavior.

After benign fine-tuning, the geometric change is highly localized: the safety Fisher selectively re-sharpens late output-side MLP modules.  
In particular, the final-layer \texttt{down\_proj} sharpness increases by \(11.2\times\) for safety, whereas the corresponding increase for utility is only \(1.2\times\).
This localized re-sharpening provides a bridge from geometry to behavior.
If refusal depends on an output-side routing path, then its high curvature means that small updates can strongly perturb this routing.  
By contrast, because the utility Fisher exhibits a disproportionately lower re-sharpening, the same updates have a much weaker effect on the knowledge-task output.  
This explains the \emph{asymmetric fragility} (\S\ref{sec:rq1}). Aligned models with near $0\%$ ASR suffer severe safety collapse (up to $85.7\%$ ASR) after only \(100\) benign examples across model families, alignment settings, and datasets, whereas the corresponding utility degradation remains comparatively limited.

The routing view also explains why broken safety is reversible
(\S\ref{sec:rq4}).  
Few safety examples can restore refusal behavior, and even a fixed refusal-style prefix can reduce ASR by \(50\%\) without any parameter updates.  
Since the prefix adds no new safety knowledge, this suggests that safety-relevant representations remain intact, while fine-tuning disrupts their mapping to refusal behavior.
Consistently, LoRA and ASAM reduce early collapse by restricting
small-data output-side drift, but their protection weakens at larger data
scales as accumulated drift overwhelms the routing geometry (\S\ref{sec:rq3}).

Together, our results suggest that early safety failure is neither caused solely by adversarially selected outliers nor by a global knowledge loss. 
Instead, safety alignment is fragile because it depends on a low-rank output routing. Benign fine-tuning re-sharpens late MLP routing modules, causing refusal failure while much of the internal representation remains intact.

\mypara{Contributions} 
We make four contributions:

\noindent$\bullet$~We provide a Fisher-geometric account of alignment fragility, showing that benign fine-tuning breaks safety through selective re-sharpening of the output-side routing subspace rather than through gradient conflict or global representation drift.

\noindent$\bullet$~Our empirical analysis shows that this fragility consistently appears across the evaluated model families, alignment conditions, and sample-selection strategies, while the preservation of intermediate safety representations helps explain its reversibility.

\noindent$\bullet$Through logit-lens analysis and cross-condition activation patching, we show that safety-relevant representations remain present after benign fine-tuning, while late-layer computation causally controls the switch between refusal and compliance, providing mechanistic evidence for the output-routing account and explaining the reversibility of safety.

\noindent$\bullet$~We show that sharpness-oriented mitigations (LoRA, ASAM) address the routing fragility at small scale but face a fundamental ceiling at larger data volumes, motivating alignment methods beyond sharpness control.

\section{Preliminaries}
\label{sec:prelim}

\subsection{Background}
\label{sec:background}
 
\mypara{Aligned models and refusal behavior}
We study a chat model with parameters $\theta \in \mathbb{R}^{n}$ and
policy $\pi_{\theta}(y \mid x)$. We write $\theta_\mathrm{safe}$ for the aligned checkpoint produced by SFT or DPO~\citep{rafailov2023direct}, which both retains general instruction-following ability and refuses harmful queries. 
We measure safety via the Attack Success Rate
\begin{equation}
\mathrm{ASR}(\theta) :=
\tfrac{1}{|\mathcal{H}|}\!\sum_{x \in \mathcal{H}}
\mathbf{1}\!\bigl[\mathrm{Judge}(\pi_{\theta}(\cdot\mid x))
                   =\text{unsafe}\bigr]
\label{eq:asr}
\end{equation}
on the HEx-PHI benchmark~\citep{qi2024fine} with an LLM-as-a-judge protocol. 
By design, $\mathrm{ASR}(\theta_\mathrm{safe})\approx 0$.
 
\mypara{Benign fine-tuning and safety collapse}
Practitioners adapt $\theta$ or $\theta_\mathrm{safe}$ to downstream tasks by minimizing
the supervised fine-tuning objective
\begin{equation}
L_{\mathrm{ft}}(\theta) =
-\mathbb{E}_{(x,y)\sim\mathcal{D}_{\mathrm{ft}}}
\bigl[\log\pi_{\theta}(y\mid x)\bigr]
\label{eq:lft}
\end{equation}
on a benign dataset $\mathcal{D}_{\mathrm{ft}}$ that contains no
harmful content and no refusal demonstrations. Yet the resulting model
$\theta_{\mathrm{ft}}$ typically exhibits sharply elevated ASR,
sometimes saturating after as few as 100 training
examples~\citep{qi2024fine,chen2024janus}.
Prior work attributes this collapse to gradient conflict with a small subset of outlier samples~\citep{he2024your,guan2025benign},
aggressive optimization hyperparameters~\citep{kim2025rethinking}, or
the curvature structure of the loss landscape~\citep{wei2024safety,zheng2024prompt,springer2026geometry,peng2024navigating}. 
This paper takes a geometry-first stance: we argue that benign fine-tuning disrupts an \emph{output-side routing mechanism} that implements refusal safety, and that this view jointly explains both the fragility and the recoverability of alignment.
 
\subsection{Related Work}
\mypara{LLM Safety Alignment} 
To prevent LLMs from generating harmful, biased, or restricted content, researchers employ various safety alignment techniques during the post-training phase. Standard practices typically involve Supervised Fine-Tuning (SFT) on curated instruction-following demonstrations \cite{wei2021finetuned, sanh2022multitask}, followed by optimization using human preferences \cite{ouyang2022training, bai2022training} with Reinforcement Learning techniques \cite{schulman2017proximal, rafailov2023direct}. While these alignment techniques successfully train models to recognize harmful intent and map them to refusal behaviors, the training outcomes are usually fragile and easy to break after further fine-tuning \cite{yang2023shadow, lermen2023lora, zhan2024removing}.

\mypara{The Fragility of Safety Alignment}
To explain why even benign fine-tuning erodes safety, prior work has explored multiple angles. A prominent line of work attributes safety collapse to gradient conflict \cite{he2024your, qi2024fine, guan2025benign,hsiung2025llm}, arguing that specific outlier benign samples carry gradients that point in conflicting directions relative to safety-relevant parameters, causing utility updates to overwrite safety parameters. Another line of research has shifted towards understanding the structural representation of safety within model weights \cite{wei2021finetuned, arditi2024refusal, li2025safety, zhao2025understanding,ponkshe2025safety}, which demonstrates that safety behavior is controlled by a very small number of neurons, layers, and directions, and therefore safety is encoded in a remarkably low-dimensional subspace. Narrow fine-tuning will easily break safety when it interferes with the shared latent dimensions \citep{giordani2025re}. Additionally, \cite{kimrethinking} examines this fragility from an optimization standpoint, suggesting that aggressive optimization hyperparameters play a massive role in destabilizing safety behavior. Compared to these lines of work, we pinpoint a distinct geometric mechanism behind safety fragility. We use Fisher-geometric analysis to localize the failure to output-side re-sharpening of a low-rank refusal-routing pathway.
This mechanism explains both the rapid collapse of refusal behavior and its reversibility.

\section{Geometry of Alignment Collapse}
\label{sec:shortcut}

LLMs lose their safety alignment after only a handful of benign fine-tuning steps~\citep{qi2024fine, he2024your, guan2025benign}. The dominant explanation invokes \emph{gradient conflict}: a small subset of fine-tuning samples carries gradients that conflict with safety-relevant parameter directions, so utility updates overwrite safety updates~\citep{he2024your, guan2025benign}. 
Under this view, safety should be preserved whenever the benign dataset is free of such conflicting outliers. 
However, our experiments (\autoref{sec:rq2}) contradict this prediction: random, gradient-top, and gradient-bottom subsets all break safety to a comparable degree. Outlier-conflict is thus a \emph{sufficient} but not a \emph{necessary} cause of the collapse we observe.

This section therefore shifts from a sample-centric explanation to a geometry-centric one.  
Rather than asking which benign samples have the most conflicting gradients, we ask how benign fine-tuning changes the \emph{local Fisher geometry}. We find that alignment compresses the model's safety-Fisher sharpness relative to the baseline, leaving the internal safety geometry flatter and more stable.  

Benign fine-tuning then induces a localized curvature drift: it does not destroy this representation, but selectively re-concentrates sharpness in late MLP modules, disrupting the \emph{routing}, i.e., the output-side mapping that determines whether safety-relevant internal representations drive refusal or are overridden by compliant generation.

\subsection{Geometric Setup}
\label{sec:shortcut_setup}

We distinguish a baseline checkpoint \(\theta_{\mathrm{base}}\) and, for each capability \(c\), two
capability-specific checkpoints: a reference model \(\theta_c\) and its benign fine-tuned version \(\theta_c^{+}\).  
We consider $c\in\{\mathrm{safe},\mathrm{util}\}$,
where \(\mathrm{safe}\) denotes refusal safety and \(\mathrm{util}\)
denotes a general-utility/knowledge capability.  
In particular, \(\theta_{\mathrm{safe}}\) is the
safety-aligned model, while \(\theta_{\mathrm{util}}\) is the
knowledge-task SFT model.  
Their post-training counterparts are \(\theta_{\mathrm{safe}}^{+}\) and \(\theta_{\mathrm{util}}^{+}\).

For a given analysis dataset
\(\mathcal D=\{(x_i,y_i)\}_{i=1}^{N}\), we estimate a block-wise
empirical Fisher for each layer-module block \(b=(\ell,m)\)\footnote{Transformer layers $\ell$ and module
types \(m\in\mathcal M\), including modules \(\texttt{k}\), \(\texttt{q}\), \(\texttt{v}\),
\(\texttt{o}\), \(\texttt{up}\), \(\texttt{down}\),
and \(\texttt{gate}\).}:
\begin{align}
\widehat F_{b}(\theta)
&=
\frac{1}{N} \sum_{i=1}^{N}
g_{b}^{(i)}(\theta)g_{b}^{(i)}(\theta)^{\!\top},
\\
g_{b}^{(i)}(\theta)
&=
\nabla_{\theta_b}
\log\pi_{\theta}(y_i\mid x_i),
\label{eq:empirical_block_fisher}
\end{align}
where \(\pi_{\theta}(y\mid x)\) denotes the conditional distribution
over output sequences induced by parameters \(\theta\).  
For safety, \(y_i\) is the refusal target; for utility, \(y_i\) is the correct task answer.

Let
$\lambda_{b,1}(\theta)
\ge
\lambda_{b,2}(\theta)
\ge
\cdots
\ge 0
$
be the eigenvalues of \(\widehat F_b(\theta)\).  We use the top eigenvalue as the block \emph{sharpness}, measuring worst-direction curvature:

\begingroup
\small
\[
A_{b}(\theta)
=
\lambda_{\max}
\left(
\widehat F_{b}(\theta)
\right)
=
\lambda_{b,1}(\theta)
=
\max_{\|v\|=1}
v^{\top}
\widehat F_{b}(\theta)
v.
\]
\endgroup


We define curvature drift between two checkpoints
\(\theta_{\mathrm{src}}\) and \(\theta_{\mathrm{tgt}}\) as the
log-ratio change in Fisher curvature statistics:
\begin{align}
    \Delta A_{b}^{\mathrm{src}\rightarrow \mathrm{tgt}}
=
\log_{10}
\frac{
A_{b}(\theta_{\mathrm{tgt}})
}{
A_{b}(\theta_{\mathrm{src}})
}.
\end{align}
Here positive drift indicates increased local curvature at the later
checkpoint, while negative drift indicates decreased local curvature.

\subsection{Where the Post-Training Drift Lives}
\label{sec:shortcut_localization}

We use the block-wise empirical Fisher on $100$ safety inputs sampled from the Align-10k post-training set (\S\ref{sec:setup}) and $100$ SciQ~\cite{SciQ} inputs, to localize safety- and utility-relevant curvature and track how it shifts across checkpoints.
For safety, we compare
\(\theta_{\mathrm{base}}\), \(\theta_{\mathrm{safe}}\), and
\(\theta_{\mathrm{safe}}^{+}\).  For utility control, we compare \(\theta_{\mathrm{base}}\), \(\theta_{\mathrm{util}}\), and
\(\theta_{\mathrm{util}}^{+}\).

\begin{figure}[ht]
    \centering
    
    \begin{subfigure}{\linewidth}
        \centering
        \includegraphics[width=\linewidth]{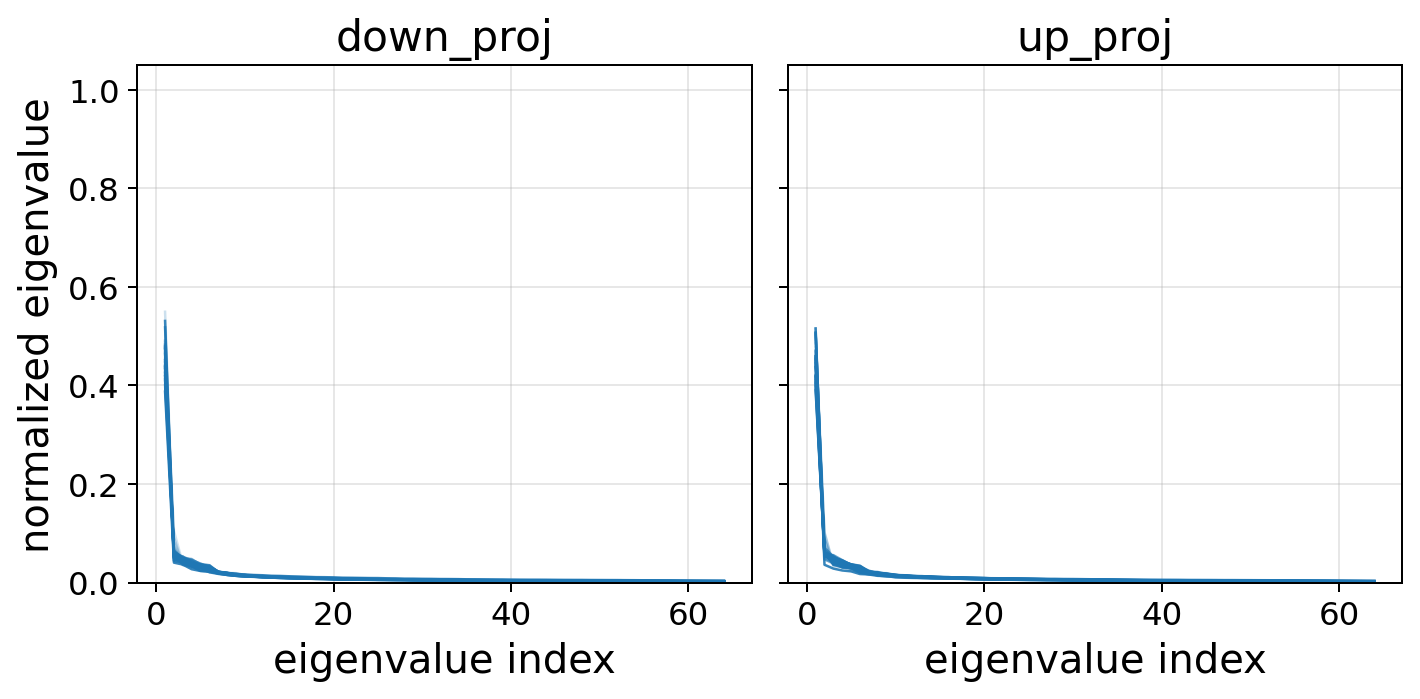}
    \end{subfigure}
    
    \vspace{-3pt} 

    \begin{subfigure}{\linewidth}
        \centering
        \includegraphics[width=\linewidth]{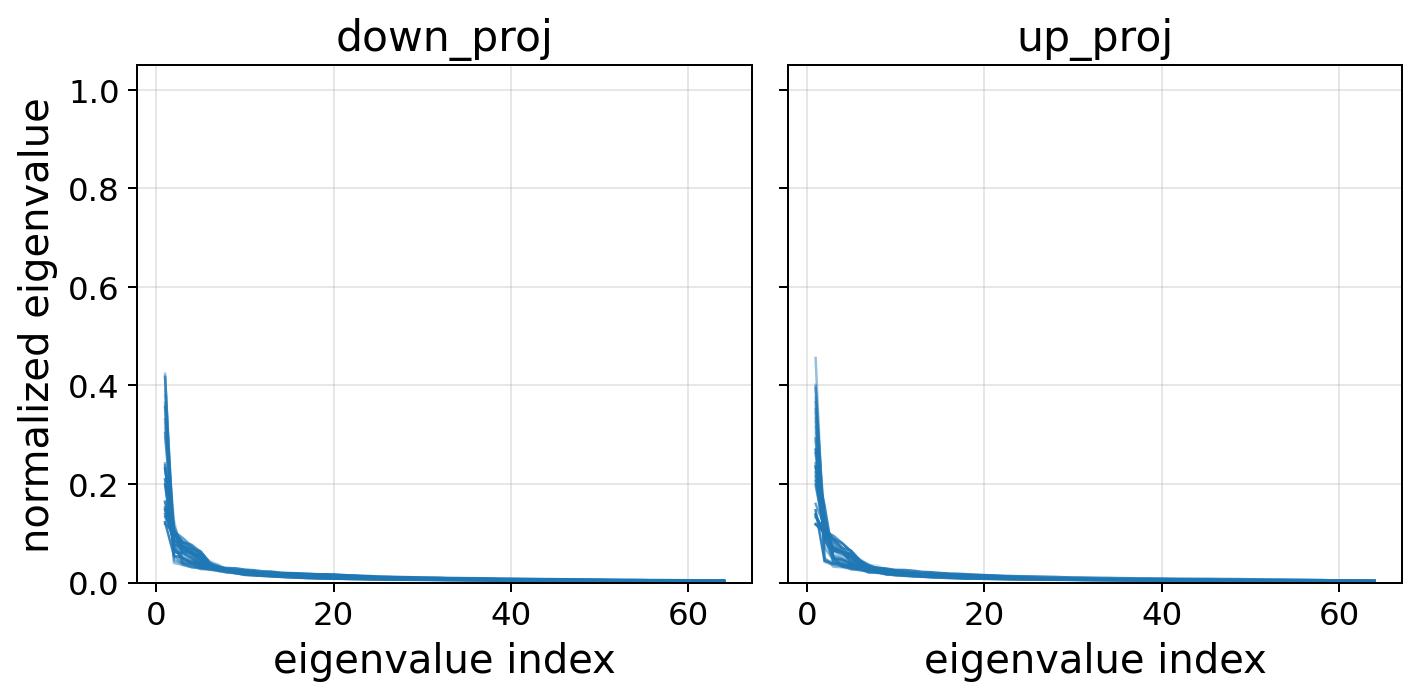}
    \end{subfigure}
    \caption{Normalized top eigenvalue decay of the Fisher on \(\theta_{\mathrm{base}}\). Individual lines represent different transformer layers. \textbf{Top}: Safety task. \textbf{Bottom}: Utility task. While both domains exhibit low-rank structure, the safety task demonstrates a substantially sharper eigenvalue decay.}
    \label{fig:fisher_decay}
\end{figure}

\mypara{Safety Fisher is low-effective-rank}
We first compare the normalized eigenvalue decay on \(\theta_{\mathrm{base}}\).  
For each
\texttt{down\_proj} and \texttt{up\_proj} block, we normalize the top-64 eigenvalues by the sum of eigenvalues, so that the comparison reflects spectral shape rather than absolute Fisher scale.
As shown in Figure~\ref{fig:fisher_decay}, both safety and utility exhibit low-rank structure.  
However, the safety spectrum decays more sharply, indicating that safety Fisher mass is more concentrated in a small number of dominant directions.

We quantify this concentration using the top-\(k\) mass ratio over all the layer-module blocks,
\[
R(k)
=
\frac{
\sum_b \sum_{j=1}^{k}\lambda_{b,j}
}{
\sum_b \sum_{j=1}^{64}\lambda_{b,j}
}.
\]
Safety has a substantially larger top-1 concentration than utility
(\(0.491\) vs. \(0.271\)), and also a larger top-5 concentration
(\(0.657\) vs. \(0.516\)).  This gap is especially pronounced in the
final layers (\autoref{tab:late_layer_concentration}).
The same pattern appears under a normalized-eigenvalue rank estimate.  We
count the number of normalized eigenvalues larger than \(1/n\), where
\(n\) is the matrix dimension, following the Kaiser criterion.  This gives
an estimated rank of \(8\) for safety and \(11\) for utility.  Thus, the
safety Fisher is more strongly concentrated than the utility Fisher, with
fewer directions accounting for a larger fraction of the measured curvature.

\begin{table}[t]
\centering
\small
\resizebox{\linewidth}{!}{
\begin{tabular}{c c c c c}
\toprule
Layer & Safety \(R(1)\) & Safety \(R(5)\) & SciQ \(R(1)\) & SciQ \(R(5)\) \\
\midrule
29 & 0.626 & 0.756 & 0.143 & 0.397 \\
30 & 0.718 & 0.846 & 0.223 & 0.473 \\
31 & 0.431 & 0.555 & 0.185 & 0.365 \\
\textbf{Avg} & 0.491 & 0.657 & 0.271 & 0.516 \\
\bottomrule
\end{tabular}
}
\caption{Layer-wise spectral concentration of the Fisher.}
\label{tab:late_layer_concentration}
\end{table}

\mypara{Alignment compresses safety Fisher}
We next examine the absolute scale of the safety Fisher.  While the
normalized spectra above show that safety is low-rank, they do not tell
us whether the corresponding directions are large or small in absolute
curvature.  We therefore compare the layer-wise top eigenvalue
\(A_b(\theta)\) across checkpoints.

\begin{figure}[ht]
    \centering
    \includegraphics[width=0.8\linewidth]{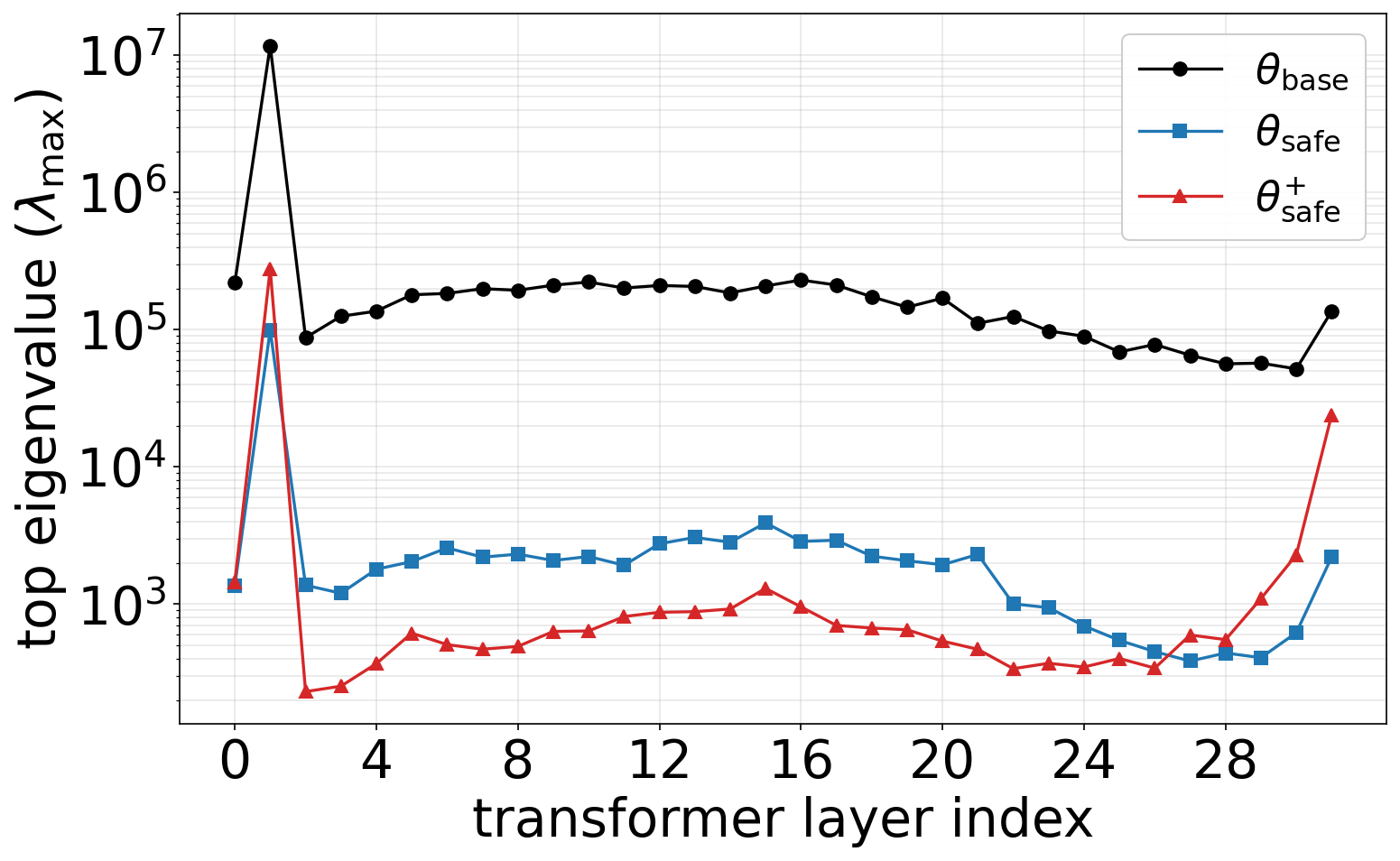}
    \caption{Layer-wise safety sharpness in \texttt{down\_proj}.}
    \label{fig:layer-wise-sharpness}
\end{figure}

As shown in \autoref{fig:layer-wise-sharpness}, alignment substantially
compresses the safety Fisher.  The baseline model
\(\theta_{\mathrm{base}}\) has large top eigenvalues across almost all
layers, whereas the aligned model \(\theta_{\mathrm{safe}}\) is lower by
roughly two orders of magnitude over most of the network.  Thus, the
aligned safety geometry is not globally sharp: after alignment, the
measured safety Fisher becomes much smaller and flatter.
This compression is not permanent.  After benign fine-tuning, \(\theta_{\mathrm{safe}}^{+}\) remains flatter than the baseline through most middle layers, but the final output-side blocks begin to re-sharpen.  
We analyze this re-concentration next.

\begin{figure}[t]
    \centering
    \begin{subfigure}{0.5\linewidth}
        \centering
        \includegraphics[width=\linewidth]{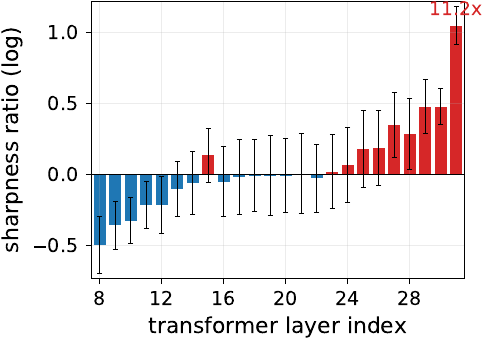}
        \caption{Safety}
        \label{fig:safety-ratio-ci}
    \end{subfigure}%
    \begin{subfigure}{0.5\linewidth}
        \centering
        \includegraphics[width=\linewidth]{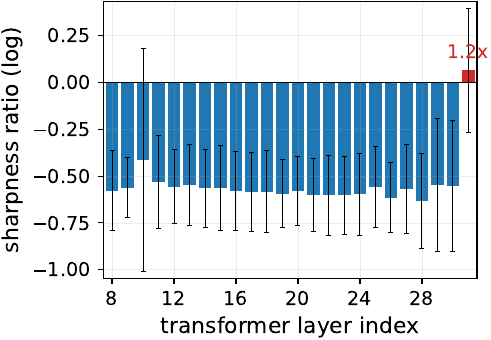}
        \caption{Utility}
        \label{fig:utility-ratio-ci}
    \end{subfigure}
    \caption{Layer-wise curvature drift in \texttt{down\_proj} after FT-100. Bars show the mean across five Fisher estimation seeds, with 95\% confidence intervals. Positive bars indicate re-sharpening, while negative bars indicate further flattening.}
    \label{fig:down-proj-ratio-comparison-ci}
\end{figure}


\mypara{Fine-tuning selectively re-sharpens safety in late MLP projections}
We next examine curvature drift from 
\(\theta_c\) to the post-training \(\theta_c^+\).  In
\autoref{fig:down-proj-ratio-comparison-ci}, we plot the layer-wise
sharpness ratio for the MLP \texttt{down\_proj} module, averaged over
five Fisher-estimation seeds with 95\% confidence intervals.
For safety, FT-100 does not increase curvature uniformly across the model. 
The top eigenvalues of middle layers continue to decrease. 
In contrast, the final layers flip to positive drift, with the last layer increasing by \(11.2\times\). 
Thus fine-tuning selectively re-sharpens the output-side MLP projection, making the safety subspace locally high-curvature and less robust.

The SciQ utility control shows a different pattern: most layers remain flatter after FT-100, 
and the only visible late-layer increase is modest, reaching about \(1.2\times\) in the final layer.  
This weaker curvature drift is consistent with the behavioral results in \autoref{tab:main_asr_utility}, where fine-tuning reduces
SciQ accuracy by only about \(10\%\), compared to the much sharper drop in refusal safety.

This contrast suggests that the safety failure is not well explained by a global loss of knowledge.  
Instead,
benign fine-tuning renders the late MLP refusal-routing pathways fragile and highly curved, while preserving utility geometry. 
Thus, the early drop in safety is best understood as a localized disruption of output-side routing.
This interpretation suggests two interventions that we return to later: flattening the optimization trajectory to reduce output-side re-sharpening (\S\ref{sec:rq3}), and directly repairing the disrupted refusal-routing path, which would explain why safety can be recovered (\S\ref{sec:rq4}).
Both interventions build directly on this localization, and motivate the next step: the curvature shift we measure lives in parameter space, telling us where fine-tuning moves the model, but not yet what the model computes when it answers a harmful prompt. 
To establish that the final layers are causally responsible for the loss of refusal, we further move from parameter-space geometry to model activations.

\subsection{From Geometry to Causal Analysis}
\label{sec:rq5}

The Fisher results show that benign fine-tuning selectively re-sharpens late
output-side \texttt{down\_proj} blocks, localizing the strongest
safety-specific change to the final layers
(\autoref{fig:down-proj-ratio-comparison-ci}). We next test whether this
localization is reflected in model activations and is causally responsible
for refusal behavior. A logit-lens read-out first traces refusal and
compliance signals across layers, giving correlational evidence on whether
the refusal signal survives benign fine-tuning; cross-condition activation
patching then tests the causal role of late-layer activations directly.

\mypara{Logit-lens read-out}
Let $x$ denote a harmful prompt and
$h_{\ell}(x)\in\mathbb{R}^{d}$ the residual-stream hidden state at the first
generation position after decoder layer $\ell$, where $\ell=0$ denotes the
embedding output and $\ell=1,\dots,L$ index the transformer layers.
We decode this intermediate state using the model's final normalization and
unembedding matrix $W_U\in\mathbb{R}^{|\mathcal{V}|\times d}$:
\begin{equation}
z_{\ell}(x)
=
W_U\,\mathrm{RMSNorm}\!\left(h_{\ell}(x)\right)
\in \mathbb{R}^{|\mathcal{V}|},
\label{eq:lens}
\end{equation}
where $\mathcal{V}$ is the vocabulary and $z_{\ell}(x)_t$ is the logit of
token $t$.

Let $\mathcal{R}\subset\mathcal{V}$ and
$\mathcal{C}\subset\mathcal{V}$ denote predefined refusal and compliance token
sets. We define the refusal--compliance margin as
\begin{equation}
m_{\ell}(x)
=
\max_{t\in\mathcal{R}} z_{\ell}(x)_t
-
\max_{t\in\mathcal{C}} z_{\ell}(x)_t .
\label{eq:margin}
\end{equation}
Thus $m_{\ell}(x)>0$ indicates that refusal dominates the read-out, and
$m_{\ell}(x)<0$ that compliance does. We report the
dataset-level mean
\[
\bar m_{\ell}
=
\frac{1}{|\mathcal{D}|}
\sum_{x\in\mathcal{D}} m_{\ell}(x),
\]
and compare $\theta_{\mathrm{safe}}$ (Align-10k) with
$\theta_{\mathrm{safe}}^{+}$, the same checkpoint after FT-100 on Alpaca.

\begin{figure}[t]
    \centering
    \includegraphics[width=\columnwidth]{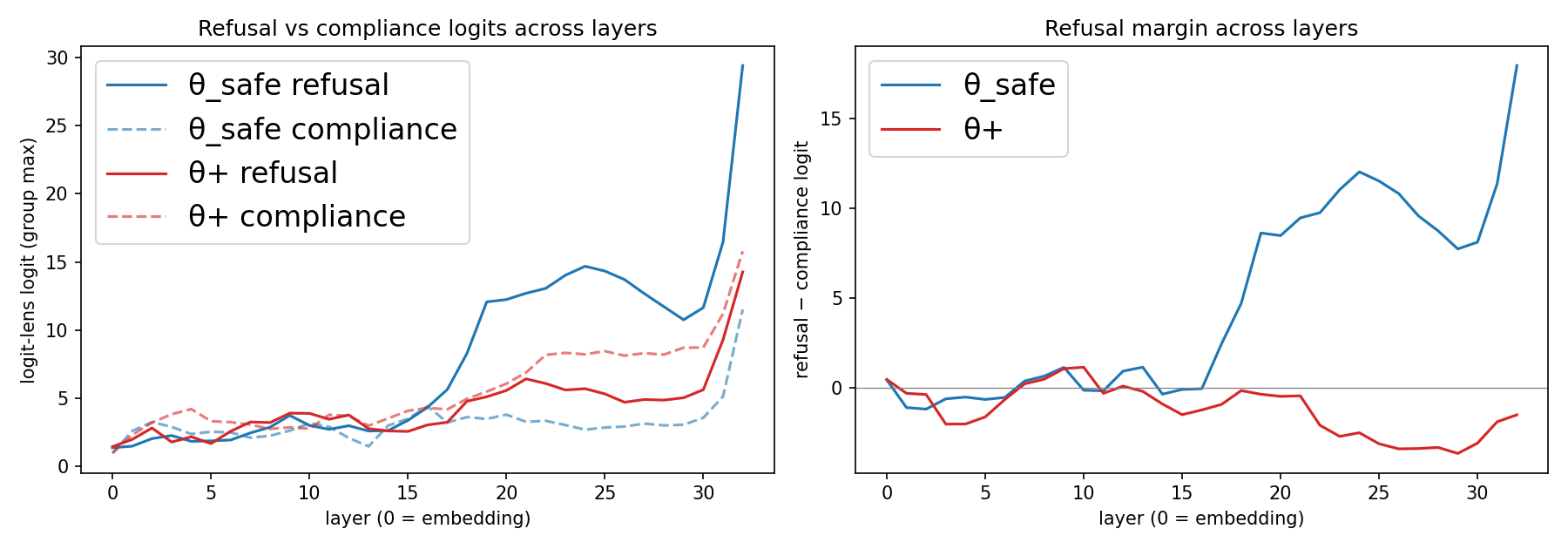}
    \caption{Logit-lens read-out at the first generation position.
    \textbf{Left}: refusal and compliance group logits across layers.
    \textbf{Right}: refusal--compliance margin $\bar m_{\ell}$.}
    \label{fig:logit-lens}
\end{figure}

\mypara{The refusal signal survives, but compliance becomes dominant}
\autoref{fig:logit-lens} shows that benign fine-tuning does not erase the
refusal signal. In $\theta_{\mathrm{safe}}$, the refusal read-out strengthens
through the late layers and eventually dominates compliance, producing a
positive margin near the output.

After benign fine-tuning, the refusal-group logit still rises through the
middle and late layers, but the compliance signal rises earlier and reaches a
higher level, preventing the margin from becoming positive. Thus,
safety-relevant information remains present, but no longer dominates the
final read-out.

If the refusal signal is merely overridden rather than erased, safety should be easy to recover, which we confirm in \S\ref{sec:rq4}.
Because the logit lens is observational, we next intervene directly on
these activations.

\begin{figure}[t]
    \centering
    \includegraphics[width=\columnwidth]{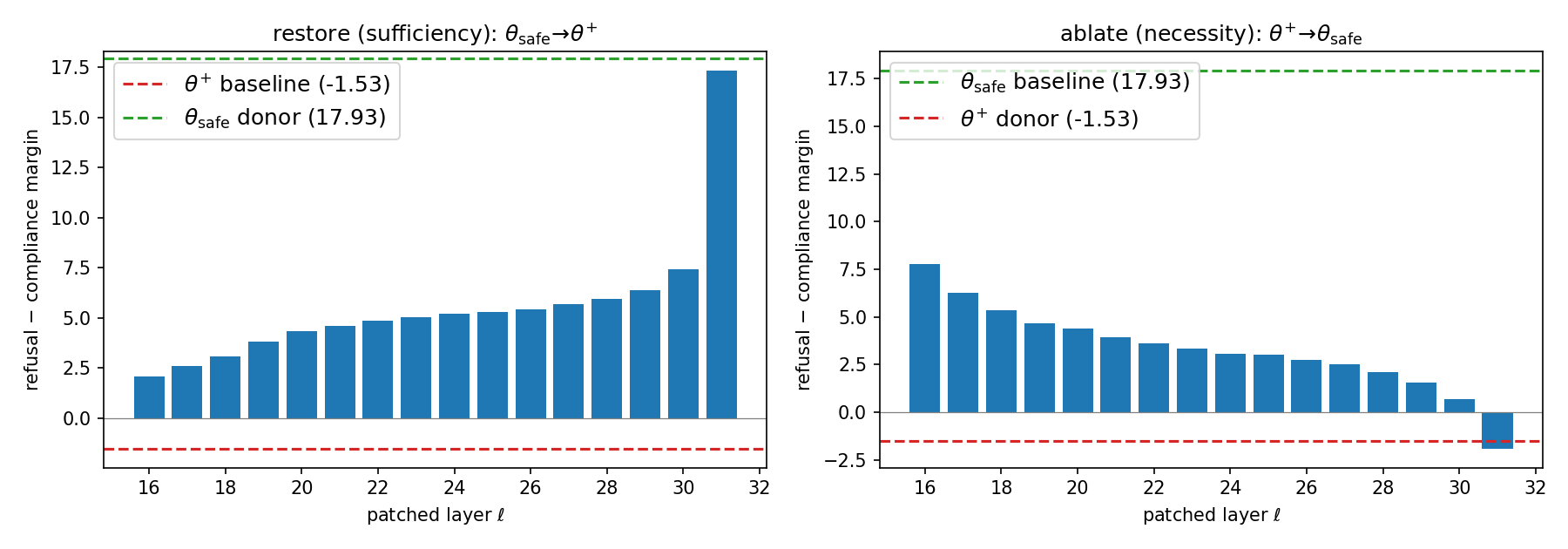}
    \caption{Cross-condition activation patching at the first generation
    position. \textbf{Left}: sufficiency. \textbf{Right}: necessity.}
    \label{fig:activation-patch}
\end{figure}

\mypara{Activation patching identifies the final layer as the dominant causal locus}
We perform cross-condition activation patching between
$\theta_{\mathrm{safe}}$ and $\theta_{\mathrm{safe}}^{+}$.
For a given layer $\ell$, we replace the base model's residual-stream state at
the first generation position with the corresponding donor activation on the
same prompt, then measure the resulting final refusal--compliance margin.

Let $\bar m^{\mathrm{base}}$ and $\bar m^{\mathrm{donor}}$ denote the
unpatched margins of the base and donor models, and
$\bar m^{(\ell)}$ the margin after patching layer $\ell$. We define
\begin{equation}
\mathrm{Effect}(\ell)
=
\frac{
\bar m^{(\ell)}-\bar m^{\mathrm{base}}
}{
\bar m^{\mathrm{donor}}-\bar m^{\mathrm{base}}
}.
\label{eq:effect}
\end{equation}
An effect of $1$ corresponds to transferring the full base-to-donor
behavioral difference with a single-layer patch.
In the \emph{restore} direction,
$\theta_{\mathrm{safe}}^{+}$ is the base and
$\theta_{\mathrm{safe}}$ is the donor, testing sufficiency for recovering
refusal. In the reverse \emph{ablate} direction, the roles are exchanged,
testing necessity for preserving refusal.

As shown in \autoref{fig:activation-patch}, the intervention effect increases
with depth and becomes dominant at the final layer. Patching the final-layer
activation from $\theta_{\mathrm{safe}}$ into
$\theta_{\mathrm{safe}}^{+}$ restores $96.9\%$ of the aligned refusal margin. In the reverse direction, the
final-layer patch eliminates the aligned refusal margin and flips it negative. These results identify the final-layer
computation as the dominant cause of the behavioral change.

\mypara{Causal evidence agrees with the Fisher geometry}
The three analyses converge on the final layer. Fisher
geometry localizes the strongest safety-specific re-sharpening to the final
\texttt{down\_proj} blocks; the logit lens shows that the refusal signal
survives but is overtaken by compliance; and activation patching establishes
that the final-layer state causally controls this behavioral switch.
Together, these results support an output-routing account of alignment
fragility: benign fine-tuning does not primarily erase safety-relevant
representations, but perturbs the late computation that determines whether
they drive refusal or are overridden by compliance.


%
%
\section{Understanding Alignment Fragility}
\label{sec:evaluation}



\subsection{Experimental Setup}
\label{sec:setup}

\mypara{Models and Alignment Conditions}
We evaluate two primary backbone families: \textsc{Llama-3.1-8B-Instruct} (hereafter \textsc{Llama3.1}) and \textsc{Qwen2.5-7B-Instruct} (hereafter \textsc{Qwen2.5}). Both models are examined across three alignment settings: (1) the default \emph{Instruct Baseline}; (2) an \emph{Align-256} variant, post-trained on 256 safety augmentation samples; and (3) an \emph{Align-10k} variant, post-trained on a 10k-scale safety dataset, where the queries consist of highly toxic and harmful prompts sourced from the PKU-SafeRLHF dataset \cite{ji2024beavertails}, and the SFT labels are explicit refusals generated by GPT-4o \cite{hurst2024gpt}.
Notably, both aligned variants achieve an initial attack success rate of $\text{ASR}=0\%$ on the HEx-PHI~\cite{qi2024fine} benchmark.

\mypara{Fine-Tuning Configurations}
We evaluate full-parameter supervised fine-tuning (SFT) against two constrained alternatives: \textbf{LoRA} (with rank $r = 32$ and scaling factor $\alpha = 64$) and \textbf{Adaptive Sharpness-Aware Minimization (ASAM)}. 
To study the effect of benign utility data, we perform a comprehensive sweep over dataset sizes $n \in \{100, 500, 1000, 5000\}$ using two widely adopted instruction-tuning datasets, \textbf{Alpaca}~\cite{alpaca} and \textbf{Dolly}~\cite{DatabricksBlog2023DollyV2}. 
All experiments are conducted with two random seeds (42 and 69) on a single NVIDIA H200 GPU.

We use the following hyperparameter settings for different training regimes:
(1) \textbf{Safety Alignment:} 5 epochs, learning rate $2 \times 10^{-5}$, batch size 20; 
(2) \textbf{ASAM:} 5 epochs, learning rate $5 \times 10^{-5}$, batch size 10, with perturbation radius $\rho = 0.05$ applied at the beginning; 
(3) \textbf{Benign Data Attack (Full SFT):} 5 epochs, learning rate $5 \times 10^{-5}$, batch size 20; 
(4) \textbf{Benign Data Attack (LoRA):} 5 epochs, learning rate $2 \times 10^{-4}$, batch size 32.
 
\mypara{Data Sampling Methods for Gradient Conflict}
To investigate the impact of gradient direction, we extract contrastive 100-sample subsets (Top/Bottom Conflict-100) from the Alpaca dataset using two distinct pipelines: (1) \textbf{Method 1 \citep{he2024your}:} Following the frameworks of~\cite{killamsetty2021grad}, we rank benign samples based on the cosine similarity between their response-token gradients and a reference adversarial gradient derived from the Pure Bad Dataset (PBD)~\cite{qi2024fine}. (2) \textbf{Method 2 \citep{guan2025benign}:} We directly deploy its data-sampling pipeline to select corresponding Top and Bottom conflict subsets.

\mypara{Evaluation Benchmarks and Metrics}
We evaluate model performance along two primary axes: \textbf{general utility} and \textbf{safety}. 
(1) General utility is multi-dimensionally measured using standard benchmarks including MMLU~\cite{mmlu}, BoolQ~\cite{clark2019boolq}, and ARC-Easy~\cite{clark2018think}, with evaluation conducted via the \texttt{lm-evaluation-harness} framework~\cite{lm_eval_harness_2024}. 
(2) Safety is primarily assessed by the Attack Success Rate (ASR) on the HEx-PHI benchmark, using an LLM-as-a-judge protocol instantiated with the GPT-4o mini API. To ensure the robustness and generalizability of our behavioral findings, we further extend our evaluation in the Appendix~\ref{sec:appendix_safety_robustness}, incorporating additional safety benchmarks (e.g., Wildchat~\cite{zhao2024wildchat} and StrongReject~\cite{souly2024strongreject}).

To further contextualize alignment fragility, we compare safety degradation against other learned capabilities. 
Specifically, we evaluate performance on two auxiliary datasets: (i) SciQ~\cite{SciQ}, measured by accuracy; and (ii) MUSE-News~\cite{shimuse}, measured by the ROUGE-based \texttt{knowmem\_r} metric on the retain set. 

\begin{table*}[t]
\centering
\scriptsize
\resizebox{\textwidth}{!}{
\begin{tabular}{ll|ccc|ccc|ccc|ccc}
\toprule
\multirow{2}{*}{\textbf{Model}} &
\multirow{2}{*}{\textbf{FT Methods}} &
\multicolumn{3}{c|}{\textbf{ASR (\%)}$\downarrow$} &
\multicolumn{3}{c|}{\textbf{MMLU (\%)}$\uparrow$} &
\multicolumn{3}{c|}{\textbf{BoolQ (\%)}$\uparrow$} &
\multicolumn{3}{c}{\textbf{ARC-E (\%)}$\uparrow$} \\
\cmidrule(lr){3-5}
\cmidrule(lr){6-8}
\cmidrule(lr){9-11}
\cmidrule(lr){12-14}
& &
\textbf{Base} & \textbf{FT-100} & \textbf{FT-5k} &
\textbf{Base} & \textbf{FT-100} & \textbf{FT-5k} &
\textbf{Base} & \textbf{FT-100} & \textbf{FT-5k} &
\textbf{Base} & \textbf{FT-100} & \textbf{FT-5k} \\
\midrule
\multicolumn{14}{c}{\textbf{Llama-3.1-8B-Instruct Model}} \\
\midrule
\multirow{3}{*}{Instruct Baseline}
& \textbf{SFT}
& 6.20 & 75.40 & 71.20
& 68.64 & 55.01 & 27.97
& 83.70 & 75.32 & 72.60
& 82.15 & 70.37 & 55.43 \\
& + LoRA
& --- & 2.40 & 44.80
& --- & 68.20 & 65.30
& --- & 85.78 & 86.21
& --- & 83.67 & 76.09 \\
& + ASAM
& --- & 63.30 & 68.50
& --- & 59.59 & 35.94
& --- & 83.15 & 69.17
& --- & 69.53 & 63.43 \\
\midrule
\multirow{3}{*}{\shortstack[l]{Align-256}}
& \textbf{SFT}
& 0.00 & 85.70 & 74.80
& 66.47 & 52.46 & 30.59
& 85.47 & 73.61 & 64.56
& 79.71 & 67.63 & 56.10 \\
& + LoRA
& --- & 24.50 & 58.80
& --- & 67.54 & 64.53
& --- & 87.03 & 85.66
& --- & 83.67 & 75.42 \\
& + ASAM
& --- & 77.60 & 67.30
& --- & 56.24 & 31.90
& --- & 77.92 & 69.97
& --- & 69.40 & 63.38 \\
\midrule
\multirow{3}{*}{\shortstack[l]{Align-10k}}
& \textbf{SFT}
& 0.00 & 59.60 & 67.60
& 63.08 & 47.20 & 27.56
& 84.74 & 69.88 & 66.45
& 72.47 & 69.95 & 57.66 \\
& + LoRA
& --- & 11.00 & 31.20
& --- & 65.33 & 63.17
& --- & 85.78 & 85.23
& --- & 82.58 & 74.37 \\
& + ASAM
& --- & 42.10 & 49.40
& --- & 53.15 & 35.80
& --- & 78.29 & 69.76
& --- & 67.85 & 64.73 \\
\midrule
\multicolumn{14}{c}{\textbf{Qwen-2.5-7B-Instruct Model}} \\
\midrule
\multirow{3}{*}{Instruct Baseline}
& \textbf{SFT}
& 7.27 & 63.60 & 57.88
& 74.25 & 69.69 & 50.58
& 85.90 & 85.23 & 77.52
& 81.65 & 74.54 & 66.29 \\
& + LoRA
& --- & 4.50 & 9.09
& --- & 74.19 & 71.65
& --- & 86.64 & 86.64
& --- & 83.92 & 79.34 \\
& + ASAM
& --- & 32.70 & 60.30
& --- & 73.29 & 61.92
& --- & 86.27 & 81.50
& --- & 79.21 & 73.48 \\
\midrule
\multirow{3}{*}{\shortstack[l]{Align-256}}
& \textbf{SFT}
& 0.00 & 72.40 & 69.39
& 73.97 & 70.91 & 49.28
& 85.38 & 85.20 & 80.21
& 80.56 & 73.70 & 64.90 \\
& + LoRA
& --- & 0.10 & 21.21
& --- & 74.13 & 71.59
& --- & 86.76 & 87.52
& --- & 83.04 & 78.91 \\
& + ASAM
& --- & 18.70 & 52.12
& --- & 72.58 & 61.54
& --- & 86.30 & 84.16
& --- & 79.25 & 72.94 \\
\midrule
\multirow{3}{*}{\shortstack[l]{Align-10k}}
& \textbf{SFT}
& 0.00 & 67.50 & 57.58
& 73.71 & 68.66 & 48.53
& 85.57 & 85.93 & 78.56
& 78.83 & 71.63 & 66.20 \\
& + LoRA
& --- & 2.50 & 16.67
& --- & 73.75 & 70.51
& --- & 86.09 & 86.94
& --- & 81.94 & 77.74 \\
& + ASAM
& --- & 24.50 & 34.24
& --- & 72.64 & 62.83
& --- & 85.69 & 83.49
& --- & 78.49 & 73.11 \\
\bottomrule
\end{tabular}
}
\caption{
Attack Success Rate (ASR) on the HEx-PHI benchmark and utility performance under benign utility fine-tuning.
FT-100 and FT-5k denote full-parameter fine-tuning on 100 and 5000 Alpaca samples, respectively.
Utility is evaluated using MMLU, BoolQ, and ARC-E.
Lower ASR indicates better safety alignment, while higher utility scores indicate better downstream task performance. All reported results are averaged over two random seeds.
}
\label{tab:main_asr_utility}
\end{table*}

\subsection{Asymmetric Fragility: Safety vs. Utility}
\label{sec:rq1}

\begin{tcolorbox}[left=1mm, right=1mm, top=0.5mm, bottom=0.5mm, arc=1mm]
\emph{Is safety degradation under benign fine-tuning a consistent phenomenon across datasets, model families, and random seeds?}
\end{tcolorbox}

Table~\ref{tab:main_asr_utility} shows that safety degradation under benign fine-tuning is pervasive. Although the aligned models (Align-256 and Align-10k) achieve $0\%$ ASR before fine-tuning, only 100 benign samples can trigger substantial safety collapse. On Alpaca with the Llama3 backbone, full-parameter fine-tuning raises the ASR of the Instruct Baseline and Align-256 to $75.4\%$ and $85.70\%$, respectively, while Align-10k reaches $59.6\%$. Table~\ref{tab:rq1_asr_summary} shows the same trend on Dolly and Qwen2.5, indicating that this fragility generalizes across datasets and model families.

\mypara{Safety vs. Utility}
We further compare safety fragility with two acquired utility capabilities (SciQ and MUSE-News) under the same 100-sample Alpaca attack. As shown in Table~\ref{tab:rq5_utility_fragility}, both model families exhibit some regression in these capabilities, but the degradation is substantially smaller than the corresponding increase in ASR. This asymmetry indicates that benign fine-tuning disproportionately disrupts safety behavior, consistent with a particularly fragile output-side routing mechanism.

\begin{table}[!htbp]
\centering
\resizebox{0.48\textwidth}{!}{
\begin{tabular}{ll c ccc}
\toprule
\textbf{FT} & \textbf{Model} & \textbf{Pre-FT} & \multicolumn{3}{c}{\textbf{Post-FT ASR}} \\
\cmidrule(lr){4-6}
\textbf{Dataset} & \textbf{Family} & \textbf{Instruct Baseline} & \textbf{Instruct Baseline} & \textbf{Align-256} & \textbf{Align-10k} \\
\midrule
\multirow{2}{*}{Alpaca} & Llama3.1  & 6.20 & 75.40  & 85.70 & 59.60  \\
                        & Qwen2.5& 7.27 & 63.60  & 72.40  & 67.50 \\
\midrule
\multirow{2}{*}{Dolly}  & Llama3.1  & 6.20 & 66.90  & 71.0  & 64.50 \\
                        & Qwen2.5 & 7.27 & 73.60  & 79.60  & 67.80  \\
\bottomrule
\end{tabular}
}
\vspace{2pt}
\begin{flushleft}
\tiny $^\dagger$ \textit{Note: Aligned models (Align-256/Align-10k) have 0\% Pre-FT ASR. }
\end{flushleft}
\caption{HEx-PHI ASR (\%) after full-parameter fine-tuning on 100 benign samples of Alpaca and Dolly.}
\label{tab:rq1_asr_summary}
\end{table}

\begin{table}[ht]
\centering
\small
\setlength{\tabcolsep}{4pt}
\renewcommand{\arraystretch}{1.05}
\resizebox{\linewidth}{!}{%
\begin{tabular}{llccc}
\toprule
\textbf{Utility Task} & \textbf{Model Family} & \textbf{Instruct Baseline} 
& \textbf{Utility SFT} 
& \textbf{Utility SFT + FT-100} \\
\midrule
\multirow{2}{*}{SciQ}
 & Llama3.1 & 92.6  & 90.9  & 77.4  \\
 & Qwen2.5    & 94.1  & 96.1  & 91.4  \\
\midrule
\multirow{2}{*}{MUSE-News}
 & Llama3.1 & 44.6 & 46.2 & 39.8 \\
 & Qwen2.5   & 32.1 & 39.4 & 37.9 \\
\bottomrule
\end{tabular}%
}
\caption{Utility-task performance across checkpoints: the original instruct baseline, the utility-SFT reference model, and its benign fine-tuned version after FT-100. Higher scores indicate better utility performance.}
\label{tab:rq5_utility_fragility}
\end{table}

\subsection{Safety Degradation: Gradient Conflict vs. Generic Drift}
\label{sec:rq2}

\begin{tcolorbox}[left=0.5mm, right=0.5mm, top=0.5mm, bottom=0.5mm, arc=1mm]
\emph{Is safety degradation driven by gradient-conflicting samples, or does it arise from generic benign fine-tuning drift?}
\end{tcolorbox}

Table~\ref{tab:gradient_asr} shows that gradient directionality modulates the severity of safety degradation but is not necessary for safety collapse.

Under Method 1, the Bottom Conflict-100 subset consistently yields lower ASR than Random-100 across all alignment conditions. For Align-256, for example, ASR decreases from $83.00\%$ to $54.20\%$, indicating that avoiding highly conflicting gradients can partially mitigate safety degradation. Method 2 exhibits the same overall trend.

Nevertheless, even the Bottom Conflict-100 subsets produce substantial degradation, with ASR exceeding $52\%$ across all alignment conditions. Reducing gradient conflict therefore attenuates the extent of degradation but does not preserve safety alignment.

Overall, gradient conflict primarily modulates the \emph{severity} of safety degradation, whereas generic benign fine-tuning drift alone is sufficient to substantially disrupt refusal behavior.

\begin{table}[htbp]
\centering
\resizebox{0.44\textwidth}{!}{%
\begin{tabular}{lcc}
\toprule
\textbf{Model} & \textbf{Bottom Conflict-100} & \textbf{Random-100} \\
\midrule
\multicolumn{3}{c}{\textbf{Method 1~\citep{he2024your}}} \\
\midrule
Instruct Baseline & 52.70 & 76.20 \\
Align-256         & 54.20 & 83.00 \\
Align-10k         & 54.55 & 63.40 \\
\midrule
\multicolumn{3}{c}{\textbf{Method 2~\citep{guan2025benign}}} \\
\midrule
Instruct Baseline & 66.67 & 76.20 \\
Align-256         & 61.82 & 83.00 \\
Align-10k         & 70.61 & 63.40 \\
\bottomrule
\end{tabular}%
}
\caption{Attack Success Rate (ASR \%) on \textsc{Llama-3.1-8B-Instruct} fine-tuned on 100-sample subsets of Alpaca selected by two gradient-based ranking methods.}
\label{tab:gradient_asr}
\end{table}

\subsection{Output-Side Sharpness Mitigation}
\label{sec:rq3}

\begin{tcolorbox}[left=0.5mm, right=0.5mm, top=0.5mm, bottom=0.5mm, arc=1mm]
\emph{Do update-restricted (LoRA) and flatness-seeking (SAM-adaptive) methods mitigate safety degradation under fine-tuning, and how does this effect scale with training data size?}
\end{tcolorbox}

To recap (\S\ref{sec:shortcut_localization}), refusal safety operates through a fragile \emph{output-side routing path}, where benign fine-tuning perturbations $\Delta\theta$ can induce sharp changes along safety-sensitive directions. \textbf{LoRA} and \textbf{ASAM} mitigate this vulnerability through complementary mechanisms: LoRA restricts updates to a low-rank subspace, limiting perturbations along safety-sensitive directions, while ASAM explicitly favors flatter local loss regions. Both therefore reduce the output-side sharpness associated with early safety collapse.

\begin{figure}[t]
\centering
\includegraphics[width=0.48\textwidth]{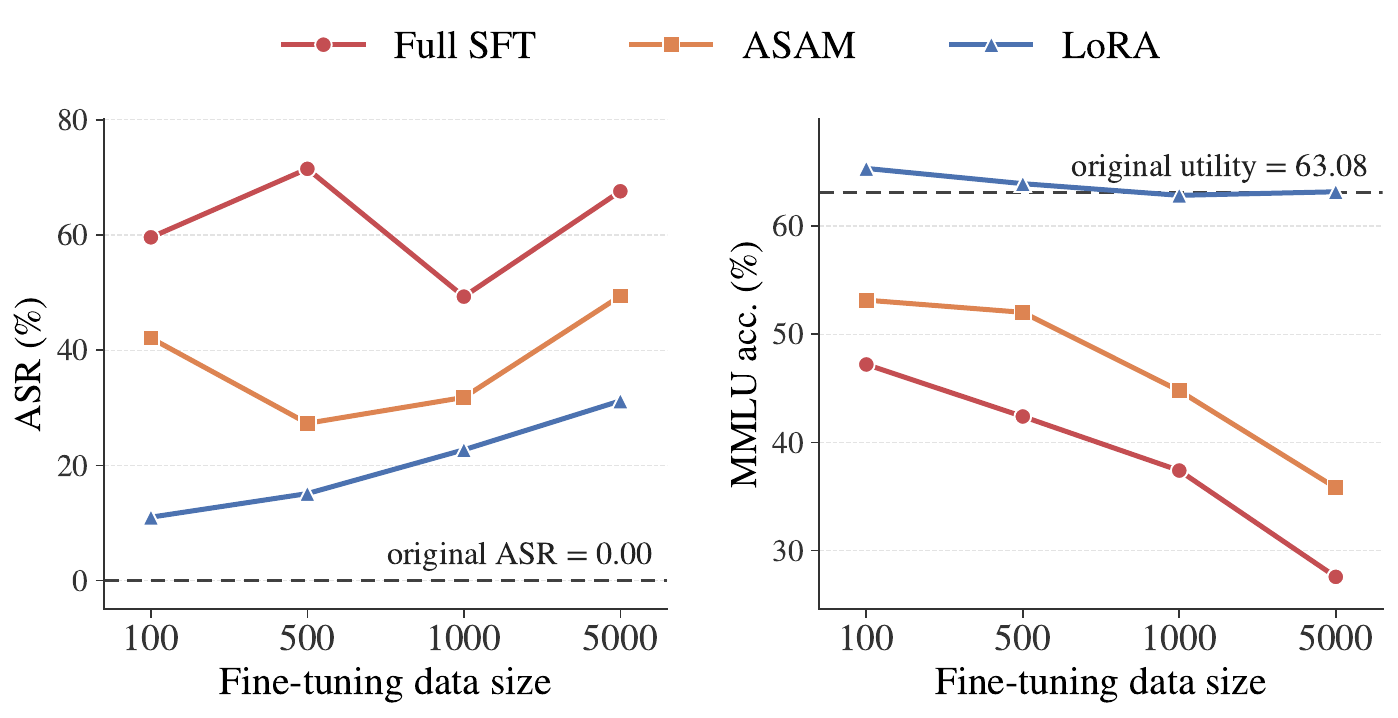}
\caption{ASR and utility (MMLU) scaling under the Align-10k model across SFT, LoRA, and ASAM.}
\label{fig:scaling-llama-saferlhf}
\end{figure}

\mypara{At 100 Samples: Sharp Drift Controlled}
As shown in Figure~\ref{fig:scaling-llama-saferlhf}, standard SFT causes an immediate increase in ASR at the 100-sample scale. Both LoRA and ASAM substantially suppress this degradation: ASAM reduces sharp local updates, while LoRA limits the magnitude of safety-relevant parameter drift. Thus, both methods better preserve the alignment routing geometry in the low-data regime.

\mypara{At 5000 Samples: Collapse Re-Emerges}
As fine-tuning scales to 5000 samples, however, these defenses become less effective. SFT and ASAM remain at high ASR, while LoRA exhibits a gradual increase as data volume grows, indicating that update restriction alone cannot prevent cumulative safety erosion. This large-scale degradation is also accompanied by declining MMLU, unlike the small-data regime, consistent with \emph{catastrophic forgetting} (CF) under extensive utility-oriented fine-tuning.

\mypara{Implications}
Results across model families and settings (Appendix~\ref{sec:scaling-trend-all-model}) support a two-regime interpretation: LoRA and ASAM mitigate early-stage collapse by reducing output-side sharpness, but do not prevent cumulative degradation at larger training scales. Robust post-training safety therefore requires mechanisms that directly protect the vulnerable output-side routing subspace during fine-tuning.

\subsection{Reversibility of Safety Alignment}
\label{sec:rq4}
\begin{tcolorbox}[left=0.5mm, right=0.5mm, top=0.5mm, bottom=0.5mm, arc=1mm]
\emph{Can safety alignment, once broken by benign fine-tuning, be efficiently recovered with minimal supervision or even without parameter updates?}
\end{tcolorbox}
Our results show that safety degradation induced by benign fine-tuning is highly reversible. Safety behavior can be restored with minimal supervision while largely preserving general utility. As shown by the performance shifts ($\Delta$) for LLaMA-3.1 in Table~\ref{tab:utility_delta_main}, benchmark performance remains broadly stable during recovery (see Appendix~\ref{sec:influence-of-recover-on-utility} for full LLaMA-3.1 and Qwen2.5 results).
Across both model families, only 10 safety examples are sufficient to reverse a 100-sample benign attack and reduce ASR to $0\%$. Even after a 5000-sample attack, 50 safety examples suffice for recovery, demonstrating strong data efficiency.
Safety behavior can also be partially reactivated \textit{without parameter updates}. Conditioning generation on a refusal prefix (e.g., ``I'm sorry'') substantially reduces ASR. For LLaMA-3.1-8B-Instruct after a 100-sample Alpaca attack, the prefix reduces ASR from $85.70\%$ to $34.85\%$ for Align-256 and from $59.60\%$ to $2.12\%$ for Align-10k.
Because a fixed prefix introduces no safety knowledge, this recovery suggests that benign fine-tuning does not erase safety-relevant representations. Instead, it disrupts the output-side routing from intent recognition to refusal generation. Recovery is also inexpensive: except for BoolQ under full SFT for Align-10k, safety restoration does not impose systemic degradation on downstream tasks. Consequently, safety compliance operates partly as a shallow, output-level mechanism---making it susceptible to benign fine-tuning, yet amenable to rapid recovery via minimal supervision or inference-time steering.

\begin{table}[htbp]
\centering
\resizebox{0.48\textwidth}{!}{
\begin{tabular}{ll ccc}
\toprule
\textbf{Model} & \textbf{FT Methods} & \textbf{$\Delta$ MMLU (\%)} & \textbf{$\Delta$ BoolQ (\%)} & \textbf{$\Delta$ ARC-E (\%)} \\
\midrule
Align-256 & SFT-5k & +2.06 & +3.51 & +1.35 \\
           & LoRA-5k     & +1.06 & +0.15 & +2.02 \\
\midrule
Align-10k & SFT-5k & +0.52 & -10.64 & -0.55 \\
& LoRA-5k     & +0.42 & -0.06 & +1.51 \\
\bottomrule
\end{tabular}
}
\caption{Changes in utility performance ($\Delta = \text{After} - \text{Before}$) after safety recovery on LLaMA-3.1.}
\label{tab:utility_delta_main}
\end{table}

\section{Conclusion}
 Refusal safety is routed through a concentrated, output-side Fisher curvature structure that benign fine-tuning can displace with as few as 100 random samples, with disproportionately smaller utility degradation. This collapse does not require adversarially selected data: gradient conflict modulates its magnitude but is not necessary. 
 Geometrically, we trace the collapse to a selective re-sharpening of Fisher curvature in late MLP blocks. LoRA and ASAM can suppress the localized drift at small data scales by reducing per-step output-side sharpness, but cannot prevent collapse at larger scales where cumulative drift overwhelms the routing geometry. 
 Taken together, these findings suggest that robust post-training safety requires moving beyond surface-level routing protection, toward alignment paradigms that distribute safety constraints more deeply.

\clearpage
\section*{Limitations}
\label{sec:limitations}

Despite the insights provided by our study on the reversibility and underlying mechanisms of safety alignment, several key limitations should be acknowledged: (1) \textit{Limited Scope of Alignment Methods}: Our investigation primarily focuses on models aligned via Supervised Fine-Tuning (SFT) and Direct Preference Optimization (DPO)~\cite{rafailov2023direct}. While these represent industry-standard and widely adopted alignment paradigms, we do not evaluate other reinforcement learning frameworks, such as Reinforcement Learning from Human Feedback (RLHF)~\cite{dai2024safe} via PPO~\cite{schulman2017proximal}, or advanced iterative preference optimization variants. Distinct alignment techniques embed safety constraints into model weights through different optimization objectives, which may exhibit varying degrees of resilience against benign fine-tuning.
(2) \textit{Model Scale and Architecture Coverage}: Our study focuses exclusively on mid-scale open-weight models (specifically the LLaMA-3.1-8B and Qwen2.5-7B families). Whether the observed alignment fragility and the hypothesized ``shallow routing'' mechanism persist in ultra-large-scale frontier models, or architectures trained with radically different synthetic data pipelines, remains an open question that warrants further empirical scrutiny. (3) \textit{Focus on English-Centric Modalities}: Due to the standard configurations of the core utility and safety benchmarks utilized in this study, our empirical findings are primarily validated on English-language corpora and instructions. Consequently, the cultural nuances, cross-lingual stability of safety alignment, and potential variations in representation routing across multilingual model spaces are left as prospective directions for future exploration.

\section*{Acknowledgments}
The research was supported by the Center for Distributed Confidential Computing (CDCC), funded by the National Science Foundation (NSF) under the grant CNS-2207031.
The research was also supported in part by Lilly Endowment, Inc, through its support for the Indiana University Pervasive Technology Institute.

\bibliography{custom}

\appendix
\label{app:data}

\section{Fisher Geometry}
\label{sec:appendix-fisher}

Let \(\theta^{*}\in\Theta\subseteq\mathbb{R}^{n}\) denote the
safety-aligned checkpoint (\(\theta_{\mathrm{safe}}\) in the notation
of \S\ref{sec:shortcut_setup}), and let \(\pi_{\theta}(y\mid x)\) be
the induced conditional distribution.
We measure local deviation from the aligned refusal behavior on harmful
prompts using the safety-deviation loss

\begingroup
\small
\begin{equation}
\mathcal L_{\mathrm{safe}}(\theta;\theta^*)
=
\mathbb E_{x\sim\mathcal D_{\mathrm{safe}}}
D_{\mathrm{KL}}
\left(
\pi_{\theta^*}(\cdot\mid x)
\;\|\;
\pi_{\theta}(\cdot\mid x)
\right)
\label{eq:safety_kl_loss}
\end{equation}
\endgroup
This loss satisfies
\(\mathcal L_{\mathrm{safe}}(\theta^*;\theta^*)=0\).
Assuming standard smoothness of \(\pi_{\theta}\), its local expansion is
\begingroup
\small
\begin{align}
\mathcal L_{\mathrm{safe}}(\theta^*+\Delta\theta;\theta^*)
&=
\tfrac{1}{2}
\Delta\theta^{\!\top}
F_{\mathrm{safe}}(\theta^*)
\Delta\theta
+
O(\|\Delta\theta\|^3),
\label{eq:fisher_local}
\\
F_{\mathrm{safe}}(\theta^*)
&=
\mathbb E_{\substack{
x\sim\mathcal D_{\mathrm{safe}}\\
y\sim\pi_{\theta^*}(\cdot\mid x)
}}
\left[
s_{\theta^*}(x,y)
s_{\theta^*}(x,y)^{\!\top}
\right],
\label{eq:fisher_def}
\\
s_{\theta^*}(x,y)
&:=
\nabla_{\theta}
\log\pi_{\theta^*}(y\mid x).
\label{eq:score_def}
\end{align}
\endgroup
Thus \(F_{\mathrm{safe}}\) is the Fisher curvature of local deviation
from the aligned safety behavior.  In experiments, we estimate a
block-wise empirical Fisher proxy on safety inputs
(harmful prompts with refusal targets; cf.\ \S\ref{sec:shortcut_setup}):
\begin{align}
\widehat F_{b}(\theta)
&=
\frac{1}{N}
\sum_{i=1}^{N}
g_{b}^{(i)}(\theta)\,g_{b}^{(i)}(\theta)^{\!\top},
\\
g_{b}^{(i)}(\theta)
&=
\nabla_{\theta_b}
\log\pi_{\theta}(y_i\mid x_i),
\end{align}
where \(b=(\ell,m)\) indexes a layer-module block.  We use
\(\widehat F_{b}(\theta)\) as a local geometric proxy for how
sensitive refusal behavior is to perturbations in block \(b\).

\mypara{Eigenvalues as directional sharpness}
The Fisher matrix \(F_{\mathrm{safe}}(\theta^*)\) is symmetric positive
semidefinite.  Let its eigenvalues be
\(\lambda_1\ge\lambda_2\ge\dots\ge\lambda_n\ge0\), with orthonormal
eigenvectors \(v_1,\dots,v_n\).  For a unit direction \(v_i\) and small
scalar \(\alpha\), \eqref{eq:fisher_local} gives
\begin{equation}
\mathcal L_{\mathrm{safe}}(\theta^*+\alpha v_i;\theta^*)
=
\tfrac12\lambda_i\alpha^2
+
O(\alpha^3).
\label{eq:directional}
\end{equation}
Thus \(\lambda_i\) is the local Fisher curvature of the safety-deviation
loss along direction \(v_i\).  The largest eigenvalue
\begin{equation}
\lambda_{\max}(F_{\mathrm{safe}}(\theta^*))
=
\lambda_1
=
\max_{\|v\|=1}
v^{\!\top}F_{\mathrm{safe}}(\theta^*)v
\label{eq:lambda_max}
\end{equation}
is the worst-case directional sharpness of the local safety geometry.
Equivalently, for any small perturbation \(\Delta\theta\),
\begin{equation}
0
\le
\tfrac12
\Delta\theta^{\!\top}
F_{\mathrm{safe}}(\theta^*)
\Delta\theta
\le
\tfrac12
\lambda_{\max}
\|\Delta\theta\|^2.
\end{equation}
A small \(\lambda_{\max}\) therefore means that the Fisher proxy is
locally flat in every unit direction, while a large \(\lambda_{\max}\)
indicates the existence of a direction in which small perturbations have
a large quadratic effect.

\mypara{The safety-relevant Fisher subspace}
For a fixed energy threshold \(\rho\in(0,1)\), define the effective
dimension
\begin{equation}
d_{\rho}
=
\min
\left\{
d:
\frac{\sum_{j=1}^{d}\lambda_j}
{\sum_{j=1}^{n}\lambda_j}
\ge
\rho
\right\}.
\label{eq:effective_dimension}
\end{equation}
We define the local safety-relevant Fisher subspace as
\begin{equation}
M_{\mathrm{safe}}(\theta^*)
=
\mathrm{span}(v_1,\dots,v_{d_{\rho}}),
\end{equation}
and let \(P_{\mathrm{safe}}\) be the orthogonal projector onto this
subspace.  For a perturbation \(\Delta\theta\), writing
\(\delta=\|P_{\mathrm{safe}}\Delta\theta\|\), the Fisher quadratic term
satisfies
\begin{equation}
\tfrac12
\Delta\theta^{\!\top}
F_{\mathrm{safe}}(\theta^*)
\Delta\theta
\ge
\tfrac12
\lambda_{d_{\rho}}
\delta^2.
\label{eq:projection_lower_bound}
\end{equation}
With the Taylor remainder included,
\begin{equation}
\mathcal L_{\mathrm{safe}}(\theta^*+\Delta\theta;\theta^*)
\ge
\tfrac12
\lambda_{d_{\rho}}
\|P_{\mathrm{safe}}\Delta\theta\|^2
-
C\|\Delta\theta\|^3
\end{equation}
for sufficiently small \(\|\Delta\theta\|\).  Hence local refusal deviation depends jointly on the sharpness of the leading Fisher directions and on how much the fine-tuning trajectory projects onto them.

\section{Additional Experimental Results}
We perform safety alignment using our fine-tuning pipeline and leverage LlamaFactory to conduct benign attack experiments and recovery experiments.
\subsection{Alignment Fragility Extends to DPO}
\label{sec:appendix_dpo}

To verify that the alignment fragility we observe is not an artifact of 
SFT-based alignment, we additionally evaluate the benign fine-tuning attack 
on a DPO-aligned model. We align the base model with DPO on our safety 
preference data using the following hyperparameters: 2 epochs, learning 
rate $1\mathrm{e}{-6}$, batch size 32, $\beta = 0.1$, and maximum sequence 
length 512. We then perform benign fine-tuning on Alpaca samples and report 
the resulting HEx-PHI ASR in Table~\ref{tab:dpo_asr} on Llama3.1.

As shown in Table~\ref{tab:dpo_asr}, the DPO-aligned model achieves $0\%$ 
ASR prior to fine-tuning, indicating strong initial safety alignment. 
However, fine-tuning on as few as $100$ benign Alpaca samples is sufficient 
to collapse this alignment, raising ASR to $66.96\%$. Scaling the 
fine-tuning data to $5{,}000$ samples yields only a marginal further 
increase to $67.88\%$. This near-identical degradation at two very different data scales mirrors the phase-transition behavior we observe for SFT-aligned models in Section~\ref{sec:rq1}, and suggests that 
the fragility we characterize is a property of the alignment surface 
itself rather than of any particular alignment algorithm.

\begin{figure*}[t]
\centering
\setlength{\abovecaptionskip}{2pt}

\begin{minipage}[t]{0.48\textwidth}
    \centering
    \includegraphics[
        width=\linewidth,
        height=0.145\textheight,
        keepaspectratio
    ]{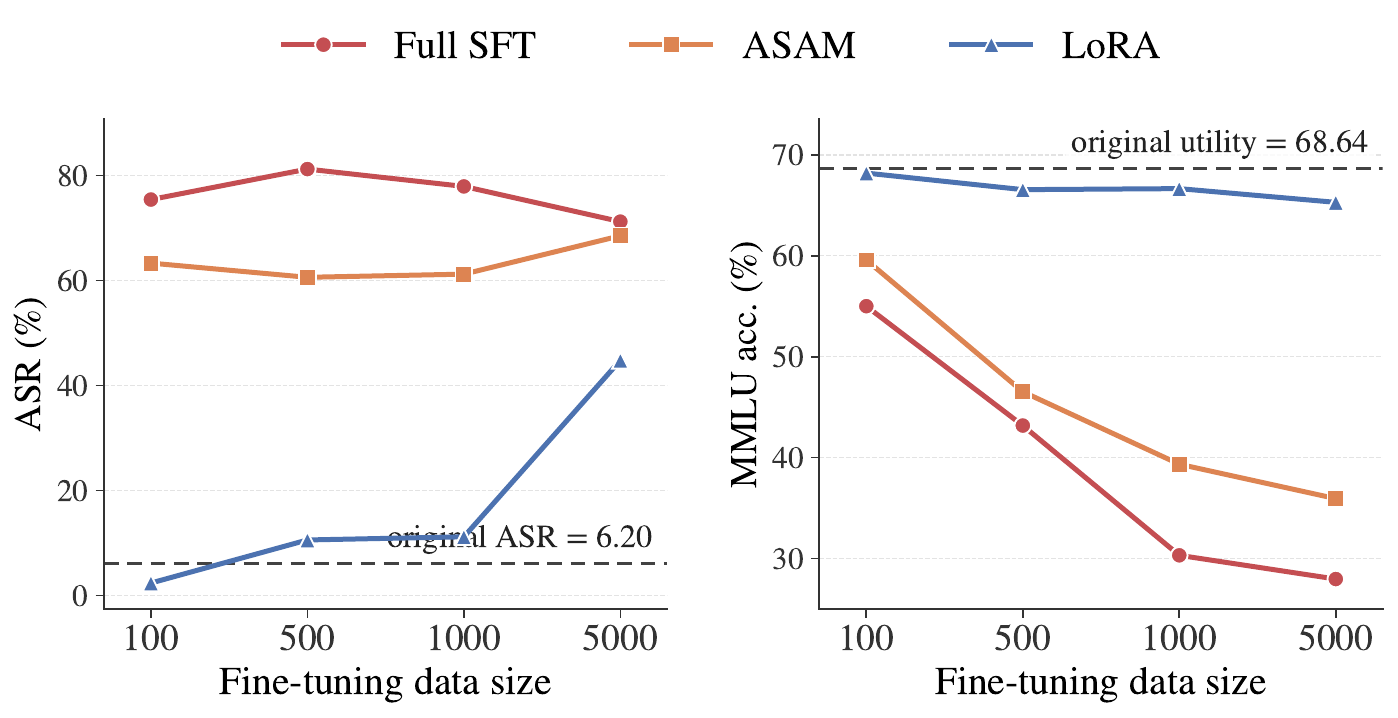}
    \caption*{(a) Llama3.1 Baseline}
\end{minipage}
\hfill
\begin{minipage}[t]{0.48\textwidth}
    \centering
    \includegraphics[
        width=\linewidth,
        height=0.145\textheight,
        keepaspectratio
    ]{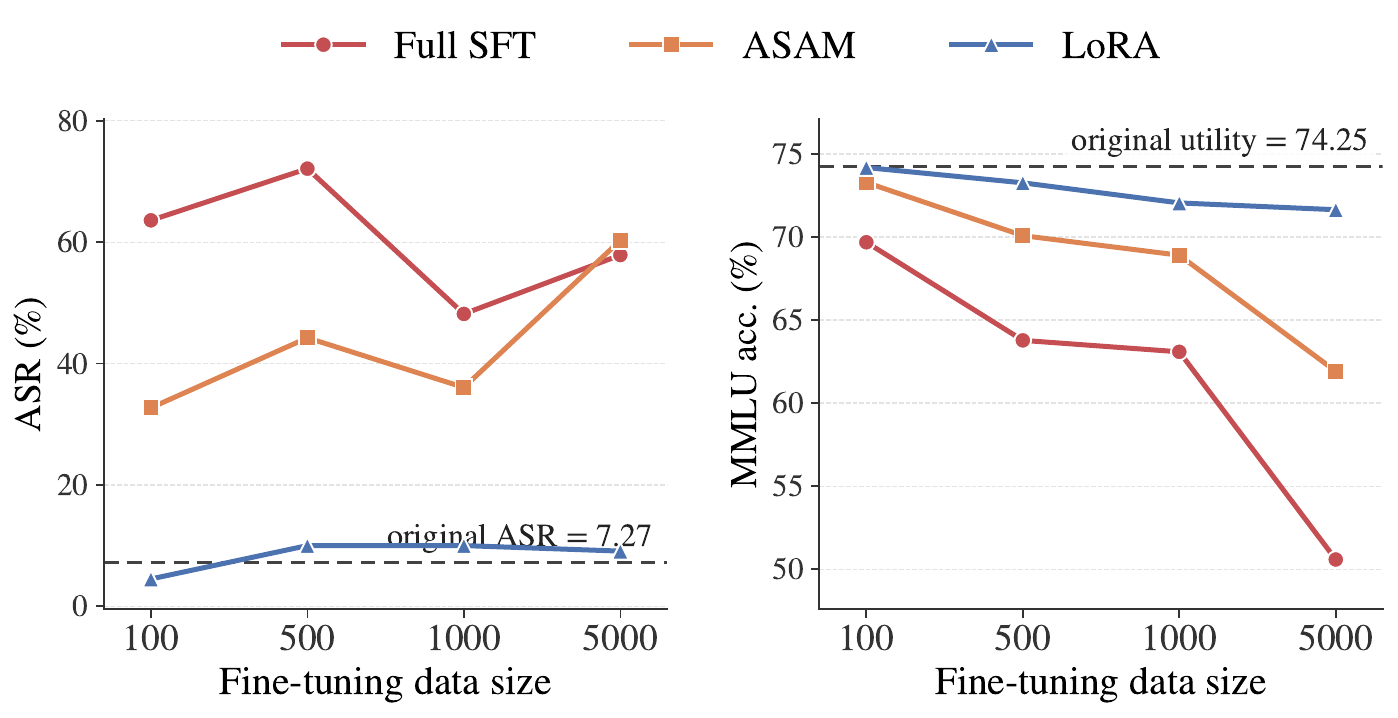}
    \caption*{(b) Qwen2.5 Baseline}
\end{minipage}

\vspace{0.15em}

\begin{minipage}[t]{0.48\textwidth}
    \centering
    \includegraphics[
        width=\linewidth,
        height=0.145\textheight,
        keepaspectratio
    ]{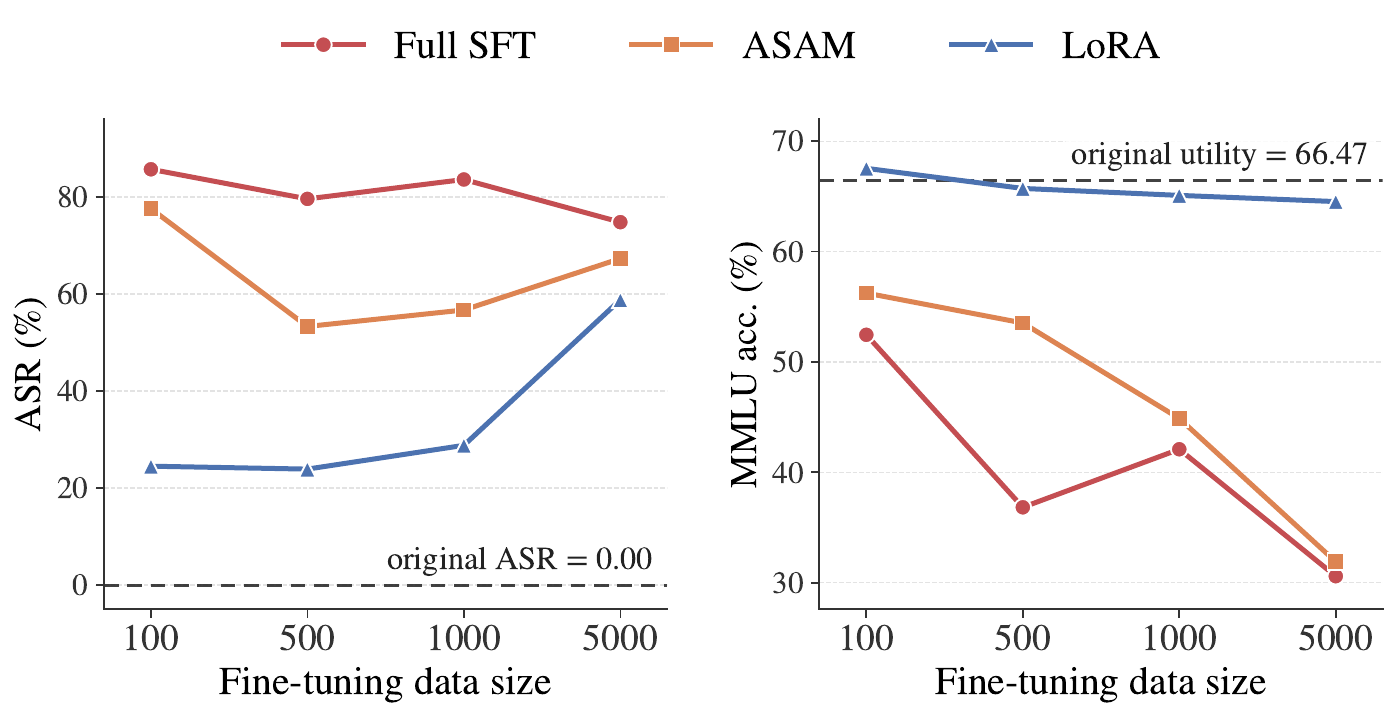}
    \caption*{(c) Llama3.1 Align-256}
\end{minipage}
\hfill
\begin{minipage}[t]{0.48\textwidth}
    \centering
    \includegraphics[
        width=\linewidth,
        height=0.145\textheight,
        keepaspectratio
    ]{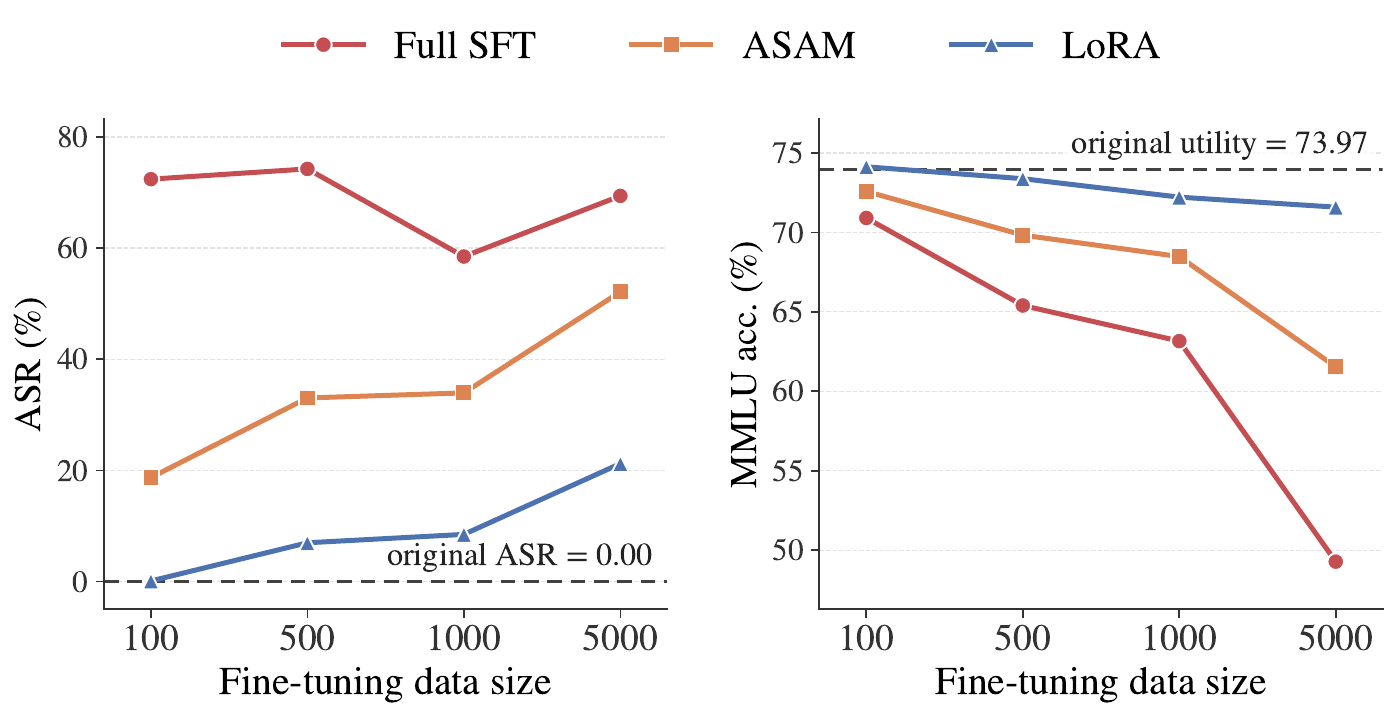}
    \caption*{(d) Qwen2.5 Align-256}
\end{minipage}

\vspace{0.15em}

\begin{minipage}[t]{0.48\textwidth}
    \centering
    \includegraphics[
        width=\linewidth,
        height=0.145\textheight,
        keepaspectratio
    ]{./figures/scaling/scaling_llama_saferlhf.pdf}
    \caption*{(e) Llama3.1 Align-10k}
\end{minipage}
\hfill
\begin{minipage}[t]{0.48\textwidth}
    \centering
    \includegraphics[
        width=\linewidth,
        height=0.145\textheight,
        keepaspectratio
    ]{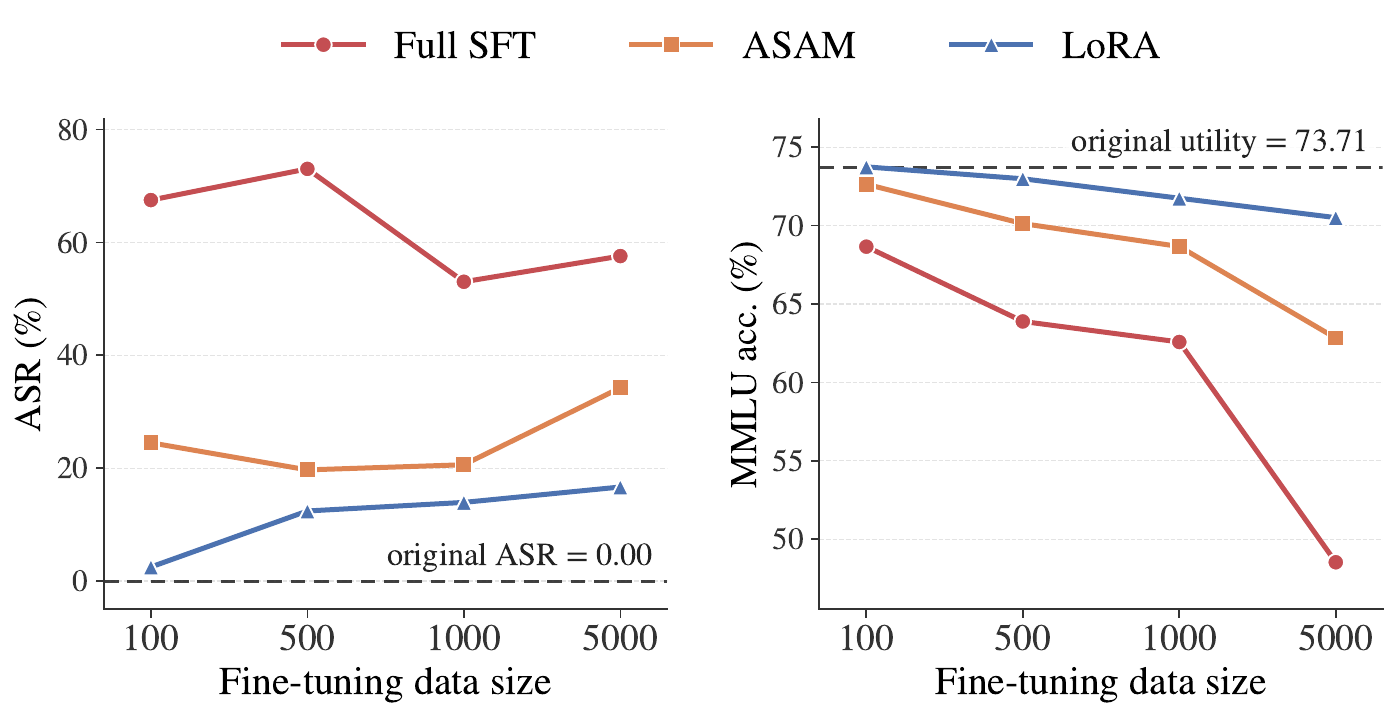}
    \caption*{(f) Qwen2.5 Align-10k}
\end{minipage}
\caption{
ASR and utility (MMLU) scaling across model families and alignment settings.
}
\label{fig:scaling-all}
\end{figure*}

\begin{table*}[!htbp]
\centering
\scriptsize
\resizebox{0.9\textwidth}{!}{
\begin{tabular}{ll l ccc ccc}
\toprule
\textbf{Model Family} & \textbf{Alignment} & \textbf{Setting} &
\multicolumn{3}{c}{\textbf{Before Recovery}} &
\multicolumn{3}{c}{\textbf{After Recovery}} \\
\cmidrule(lr){4-6}
\cmidrule(lr){7-9}
& & &
\textbf{MMLU} & \textbf{BoolQ} & \textbf{ARC-E} &
\textbf{MMLU} & \textbf{BoolQ} & \textbf{ARC-E} \\
\midrule

Llama3.1& Align-256 & SFT-5k
& 30.59 & 64.56 & 56.10
& 32.65 & 68.07 & 57.45 \\

Llama3.1& Align-10k & SFT-5k
& 27.56 & 66.45 & 57.66
& 28.08 & 55.81 & 57.11 \\

Llama3.1& Align-256 & LoRA-5k
& 64.53 & 85.66 & 75.42
& 65.59 & 85.81 & 77.44 \\

LlaMA3.1& Align-10k & LoRA-5k
& 63.17 & 85.23 & 74.37
& 63.59 & 85.17 & 75.88 \\

\midrule

Qwen2.5 & Align-256 & SFT-5k
& 49.28 & 80.21 & 64.90
& 53.59 & 81.04 & 68.06 \\

Qwen2.5 & Align-10k & SFT-5k
& 48.53 & 78.56 & 66.20
& 50.65 & 65.54 & 63.09 \\

Qwen2.5 & Align-256 & LoRA-5k
& 71.59 & 87.52 & 78.91
& 72.55 & 87.49 & 80.09 \\

Qwen2.5 & Align-10k & LoRA-5k
& 70.51 & 86.94 & 77.74
& 71.40 & 86.27 & 77.40 \\

\bottomrule
\end{tabular}
}
\caption{Granular absolute utility performance evaluation across core benchmarks before and after safety recovery execution. The recovery is accomplished via 50 dedicated safety alignment instances following initial 5000-sample Alpaca benign fine-tuning disruptions.}
\label{tab:utility_after_recovery}
\end{table*}

\begin{table}[!htbp]
\centering
\resizebox{0.9\linewidth}{!}{
\begin{tabular}{lc}
\toprule
\textbf{Condition} & \textbf{ASR (\%)} \\
\midrule
DPO-aligned              & \phantom{0}0.00 \\
\, + Alpaca fine-tuning (100 samples)     & 66.96 \\
\, + Alpaca fine-tuning (5000 samples) & 67.88 \\
\bottomrule
\end{tabular}
}
\caption{HEx-PHI ASR (\%) of the DPO-aligned model before and after benign fine-tuning on the Alpaca subset.}
\label{tab:dpo_asr}  
\end{table}

\subsection{ASR and Utility Scaling Across All Models}
\label{sec:scaling-trend-all-model}
We present the scaling trends for the remaining model variants within the Llama3.1 and Qwen2.5 families in Figure~\ref{fig:scaling-all}. Across all evaluated architectures, a highly consistent behavioral pattern emerges: localized, small-scale fine-tuning precipitously intensifies the Attack Success Rate (ASR) while leaving downstream utility largely intact. Conversely, extending the fine-tuning to a larger scale induces a much more pronounced degradation in the models' general capabilities.

\subsection{Utility Performance Profiles Surrounding Safety Recovery}
\label{sec:influence-of-recover-on-utility}

In this section, we present the comprehensive absolute evaluation scores for both the Llama3.1 and Qwen2.5 model families across core utility benchmarks (MMLU, BoolQ, and ARC-E). While Section~\ref{sec:rq4} introduces the streamlined performance deltas ($\Delta$) for presentation, Table~\ref{tab:utility_after_recovery} encapsulates the granular baseline utility profiles evaluated immediately before and after the safety remediation phase (which utilizes 50 safety alignment examples following a 5,000-sample Alpaca benign fine-tuning attack).

Across both model architectures and alignment depths (\textit{Align-256} vs. \textit{Align-10k}), the absolute capability fluctuations induced by the subsequent safety recovery process remain remarkably minimal. For instance, under the parameter-efficient LoRA setup, Qwen2.5 (Align-256) maintains steady scores across the board, shifting marginally from 71.59 to 72.55 on MMLU and from 78.91 to 80.09 on ARC-E. This strict bound on utility drift indicates that low-resource safety restoration does not require a trade-off with foundational reasoning capacities.

\begin{table}[htbp]
\centering
\small
\begin{tabular}{lcc}
\toprule
\textbf{Setting} & \textbf{Wildchat} & \textbf{StrongReject} \\
\midrule
Instruct Baseline & 16.0 & 44.83 \\
Align-10k & 64.2 & 90.48 \\
Align-10k + FT-100 & 27.0 & 46.99 \\
\bottomrule
\end{tabular}
\caption{Safety alignment evaluations on Wildchat and StrongReject, the latter reported as $1-\text{ASR}$. The auxiliary SFT recovery phase is conducted with a batch size of 8, trained for 5 epochs with a learning rate of $1\times 10^{-5}$.}
\label{tab:safety_eval_subset}
\end{table}

\subsection{Safety Evaluations Across Diverse Benchmarks}
\label{sec:appendix_safety_robustness}

To demonstrate that our observations regarding safety degradation are not specific to a particular evaluation protocol, we conduct a robustness check on two widely recognized external safety benchmarks: \textbf{Wildchat} and \textbf{StrongReject}. This cross-benchmark evaluation helps mitigate potential dataset-specific biases and assess the generalizability of our findings.

We evaluate the model across three distinct stages: the initial \textit{Instruct Baseline}, the safety-aligned model (\textit{Align-10k}), and the same aligned model after a 100-sample benign fine-tuning attack (\textit{Align-10k + FT-100}). To capture behavioral changes across benchmarks, we employ dataset-specific metrics. For Wildchat, we use the standard safety score from its official evaluation protocol, where a higher value indicates stronger safety alignment. For StrongReject, we report $1-\text{ASR}$ (Attack Success Rate), where a higher value similarly indicates a higher rate of successful refusal of harmful prompts.

As summarized in Table~\ref{tab:safety_eval_subset}, the results on both Wildchat and StrongReject are consistent with our primary findings. Safety alignment substantially improves performance over the Instruct Baseline on both benchmarks. However, subsequent benign fine-tuning on only 100 samples sharply erodes these gains, bringing safety performance closer to the baseline level. This consistent degradation across distinct evaluation protocols provides additional evidence that the vulnerability of safety alignment to benign fine-tuning is not specific to HEx-PHI or its evaluation procedure.

\end{document}